\documentclass[12pt]{article}
\usepackage{a4wide}
\usepackage{amssymb}
\usepackage{amsmath}
\usepackage{graphicx}
\usepackage{mathdots}
\usepackage{slashed}
\usepackage{verbatim}
\usepackage{xcolor}

\usepackage{hyperref}

\usepackage{IEEEtrantools}

\def\makeatletter{\catcode`\@=11}% 11:letter
\makeatletter
\def\mathbox#1{\hbox{$\m@th#1$}}%
\def\math@ccstyles#1#2#3#4#5#6#7{{\leavevmode
      \setbox0\mathbox{#6#7}%
      \setbox2\mathbox{#4#5}%
      \dimen@ #3%
      \baselineskip\z@\lineskiplimit#1\lineskip\z@
      \vbox{\ialign{##\crcr
             \hfil \kern #2\box2 \hfil\crcr
             \noalign{\kern\dimen@}%
             \hfil\box0\hfil\crcr}}}}
\def\mathaccstyles{\math@ccstyles\maxdimen}
\def\maththroughstyles{\math@ccstyles{-\maxdimen}}
\def\unity%
 {\maththroughstyles{.45\ht0}\z@\displaystyle {\mathchar"006C}\displaystyle 1}
 
\begin{document}

\begin{titlepage}

\vskip 2cm

\begin{center}
{\Large \bfseries
Supersymmetrizing 5d instanton operators
}

\vskip 1.2cm

Diego Rodriguez-Gomez,
Johannes Schmude\textsuperscript{1}

\bigskip
\bigskip

\begin{tabular}{c}
Department of Physics, Universidad de Oviedo, \\
%Facultad de Ciencias,\\
Avda.~Calvo Sotelo 18, 33007, Oviedo, Spain
\end{tabular}

\vskip 1.5cm

\textbf{Abstract}
\end{center}

\medskip
\noindent
We construct a supersymmetric version of instanton operators in five-dimensional Yang-Mills theories. This is possible by considering a five-dimensional generalization of the familiar four-dimensional topologically twisted theory, where the gauge configurations corresponding to instanton operators are supersymmetric.
\bigskip
\vfill

\footnotetext[1]{d.rodriguez.gomez@uniovi.es, schmudejohannes@uniovi.es}

\end{titlepage}

\setcounter{tocdepth}{2}

\section{Introduction and conclusions}

Five-dimensional Yang-Mills theories are naively non-renormalizable. Therefore one would be tempted to conclude that they would not generically exist as microscopic theories. Nevertheless, on the contrary to this intuition, over the recent past it has become manifest that, at least for supersymmetric theories, five-dimensional Yang-Mills theories can be at fixed points \cite{Seiberg:1996bd}, which are indeed ubiquitous.\footnote{This is easily seen upon engineering the gauge theory as a $pq$-web in string theory. See \textit{e.g.}\cite{Bergman:2013aca,Bergman:2014kza} for recent accounts.} Moreover, in many cases, 5d Yang-Mills theories exhibit amusing properties such as enhanced global symmetries. This is because in 5d there is a topologically conserved current $J\sim \star \, (F\wedge F)$ under which instanton-like solitons are electrically charged. These can combine with perturbative modes in such a way that global symmetries are enhanced. The enhanced symmetry can be both a flavor-like symmetry \cite{Seiberg:1996bd} or a Lorentz-like symmetry. The latter case is believed to happen in the maximally supersymmetric 5d Yang-Mills theory, whose instanton sector is believed to act as the KK tower which completes the theory by uplifting it to the $(2,\,0)$ 6d SCFT \cite{Lambert:2010iw}. Recently, progress towards the understanding of the underlying mechanism for these enhancements has been made in \cite{Tachikawa:2015mha} (see also \cite{Bashkirov:2012re}) by studying the quantization of gaugino zero modes --- hence directly related to the supersymmetry of the configuration --- in the instanton background.

It is natural to think of the instanton-like solitons as created by some local operator inserting the soliton at a point in spacetime and imposing certain boundary conditions on the fields \cite{Lambert:2014jna}. The corresponding classical configuration has been described for the case of $SU(2)$ gauge theories a long time ago by Yang in \cite{Yang:1977qv}. In general, we consider 5d Yang-Mills theory on $\mathbb{R}^5$, for which we choose spherical coordinates

\begin{equation}
\label{eq_polar_coords}
ds^2=dr^2+r^2\,d\Omega_5^2\, .
\end{equation}
Then, the Yang monopole \cite{Yang:1977qv} is a gauge configuration with non-vanishing

\begin{equation}
\label{eq:instanton_number}
I=\frac{1}{8\,\pi^2}\,{\rm Tr}\,\int_{S^4}\,F\wedge F\, .
\end{equation}
The simplest way to achieve this is by imposing $F_{r\mu}=0$ and $F=\pm \star_{S^4}\,F$, where $\star_{S^4}$ is the Hodge-dual operator with respect to the $S^4$ part of the metric. In appendix \ref{explicit_Yang} we review the original construction by Yang.

Since we are interested in supersymmetric Yang-Mills theories, it is natural to ask wether this configuration preserves any supersymmetry. Very recently, it has been argued in \cite{Lambert:2014jna} that the Yang monopole breaks all supersymmetries. In this note however we argue that it is possible to find other embeddings of the Yang monopole into a five-dimensional gauge theory in such a way that supersymmetry is preserved. This is possible by twisting the supersymmetry transformation with a non-trivial $SU(2)_R$ connection. Essentially this can be thought of as an uplifted version of the topological twist in four dimensions. The supersymmetry spinors are then conformal Killing spinors chiral on the $S^4$, thus allowing for supersymmetric Yang monopoles. It turns out that the $SU(2)_R$ bundle in question is actually a Yang monopole in itself --- yet for the $SU(2)_R$ background, not to be confused with the dynamical gauge field. Its second Chern class is non-trivial and it might be more appropriate to think of the theory as defined on $\mathbb{R}^5 \setminus \{ 0 \}$. Similarly, the set-up can be thought as a conformal transformation of the gauge theory on $\mathbb{R}\times S^4$ into $\mathbb{R}^5$, where, as we will see below, the conformal Killing vector $r\,\partial_r$ plays a crucial role.\footnote{Indeed, it is straightforward to find (covariantly) constant Killing spinors for the theory on $\mathbb{R}\times S^4$ upon turning on an $SU(2)_R$ gauge field that vanishes along $\mathbb{R}$.
}

It has been often the case that the topological twist allows to perform computations which carry over to the physical theory. It remains for the future to explore wether this supersymmetric version can offer insight into the dynamics of five-dimensional gauge theories, in particular on the phenomenon of enhanced symmetries, both of Lorentz-type or flavor-like.

\section{Supersymmetric Yang monopoles in five-dim.~gauge theories}

Let us consider a supersymmetric five-dimensional gauge theories on $\mathbb{R}^5$. Following \cite{Festuccia:2011ws} one can construct supersymmetric Lagrangians on arbitrary curved manifolds $M$ by coupling the theory to a suitable supergravity. In the so-called rigid limit the gravitational dynamics are frozen in such a way that we are automatically left with a supersymmetric version of the gauge theory on $M$. Background parameters in the supersymmetric gauge theory correspond to the various bosonic fields in the Weyl multiplet. They are fixed by imposing the vanishing of the gravitino and dilatino variations. Since we are concerned with a theory on flat space, that is, $M=\mathbb{R}^5$, following this approach might at first seem far too cumbersome. Yet as it will become clear below, it will allow us to find a supersymmetric embedding of the Yang monopole.

The five dimensional supergravity theory including an $SU(2)_R$ gauge field that comes to mind first is maybe the $\mathcal{N}=1$ theory of \cite{Kugo:2000af,Zucker:1999ej}. Riemannian manifolds admitting solutions to the relevant supersymmetry equations were studied by Imamura and Matsuno \cite{Imamura:2014ima} who found that these geometries always include a non-vanishing Killing vector field $\kappa$. The spinors are then essentially chiral with respect to this vector; $\slashed{\kappa} \epsilon^i = \epsilon^i$. As we will show shortly however $\kappa = r\, \partial_r$ arises naturally in the context of the Yang monopole. Since $r\, \partial_r$ is not Killing but only conformal Killing, we need to turn to a different supergravity theory. With this in mind, we will couple the gauge theory to the conformal $\mathcal{N}=2$ supergravity of \cite{Bergshoeff:2001hc, Fujita:2001kv}.\footnote{
  For a summary, see also appendix $B$ in \cite{Bergshoeff:2004kh}.
}
For simplicity, we will consider the case of a pure gauge theory. The vector multiplet contains the gauge field $A_{\mu}$, a real scalar $M$ and a triplet of auxiliary scalars $Y_{ij}$, both in the adjoint representation, as well as a symplectic Majorana doublet of gauginos $\Omega^i$. Further aspects of the theory are summarized in appendix \ref{Dirac_algebra}.

Let us now turn to the question whether the Yang monopole preserves any supersymmetry. Upon setting the gauginos as well as the $M$ and $Y_{ij}$ scalar to zero, the only non-trivial supersymmetry variation is \cite{Bergshoeff:2001hc,Fujita:2001kv}

\begin{equation}
\delta\Omega^i=-\frac{1}{4}\,\slashed{F}\,\epsilon^i\, .
\end{equation}
Since the Yang monopole configuration is such that $F_{r\mu}=0$ and $F=\pm \star_{S^4}F$, one finds that the potentially preserved supersymmetry will satisfy

\begin{equation}
\label{SUSY_condition}
\Gamma_5\,\epsilon^i=\pm\, \epsilon^i\,.
\end{equation}
where $\Gamma_5=\Gamma_1\,\Gamma_2\,\Gamma_3\,\Gamma_4$ is the chirality projector on the $S^4$ in tangent space indices. It coincides with the Dirac matrix along the fifth direction which we take to be $r$. As we mentioned in the previous paragraphs, it is this observations that let us reject the $\mathcal{N}=1$ theory of \cite{Kugo:2000af,Zucker:1999ej}. It follows from equation \eqref{SUSY_condition} that any potential supersymmetry spinor should be (anti-) chiral on the $S^4$. Moreover, we need to impose the vanishing of the gravitino and dilatino variations of \cite{Bergshoeff:2001hc,Fujita:2001kv} which leads to further constraints on the spinor. The Weyl multiplet contains an antisymmetric tensor $T$ which we set to zero. We are thus left with the constraints

\begin{eqnarray}
\label{gravitino_eq}
&&\mathcal{D}_{\mu}\epsilon^i-{\rm i} \,\gamma_{\mu}\,\eta^i=0\, , \\ && \label{chi_eq}
\frac{1}{4}\epsilon^i\,D+\frac{1}{64}\,\slashed{R}^i_{\phantom{i}j}\,\epsilon^j=0\, .
\end{eqnarray}
The covariant derivative including the $SU(2)_R$ gauge field $A$ and its field strength $R$ are defined as

\begin{equation}
\mathcal{D}_{\mu}\epsilon^i=\nabla_{\mu}\epsilon^i+(V_{\mu})^i_{\phantom{i}j}\,\epsilon^j\, , \qquad 
R_{mni}^{\phantom{mni}j} = \partial_m V_{ni}^{\phantom{ni}j} - \partial_n V_{mi}^{\phantom{mi}j} - V_{mi}^{\phantom{ni}k}V_{nk}^{\phantom{nk}j} + V_{ni}^{\phantom{ni}k} V_{mk}^{\phantom{mi}j}.
\end{equation}

By contracting \eqref{gravitino_eq} with $\gamma^{\mu}$ we find that $\eta^i$ is given in terms of $\epsilon^i$ as $\eta^i=-\frac{{\rm i}}{5}\,\slashed{\mathcal{D}}\epsilon^i$. Substituting this back into \eqref{gravitino_eq}, we find that $\epsilon^i$ is determined by the conformal Killing spinor equation

\begin{equation}
\label{CKE}
\mathcal{D}_{\mu}\epsilon^i-\frac{1}{5}\,\gamma_{\mu}\,\slashed{\mathcal{D}}\epsilon^i=0\, .
\end{equation}
Hence our task will be to find solutions to \eqref{CKE} satisfying $\Gamma_5\epsilon^i=\pm \epsilon^i$.

\subsection{The physical theory: no SUSY Yang monopoles}

Let us first consider setting to zero all background fields in the Weyl multiplet other than the metric. In the rigid limit this gives rise to the standard supersymmetric gauge theory on $\mathbb{R}^5$, to which we will refer as the physical theory.

Since after all our theory is on $\mathbb{R}^5$, in the absence of background fields it is natural to expect solutions corresponding to covariantly constant spinors. Introducing polar coordinates for the $S^4$ as

\begin{equation}
d\Omega_4^2=d\theta_1^2+\sin^2\theta_1(d\theta_2^2+\sin^2\theta_2(d\theta_3^2+\sin^2\theta_3\,d\theta_4^2))\, ,
\end{equation}
it is straightforward to see that the covariantly constant spinors are   

\begin{equation}
\label{epsilon_q}
\epsilon_q=e^{\frac{\theta_1}{2}\,\Gamma_{51}}\,e^{\frac{\theta_2}{2}\,\Gamma_{12}}\,e^{\frac{\theta_3}{2}\,\Gamma_{23}}\,e^{\frac{\theta_4}{2}\,\Gamma_{34}}\,\epsilon_q^0
\end{equation}
with $\epsilon^0_q$ being a constant 4-complex-component spinor.\footnote{One can find another class of solutions of the full conformal Killing spinor equations. Here, the spinors are of the form

\begin{equation}
\label{epsilon_s}
 \epsilon_s=r\,\Gamma_5\,e^{\frac{\theta_1}{2}\,\Gamma_{51}}\,e^{\frac{\theta_2}{2}\,\Gamma_{12}}\,e^{\frac{\theta_3}{2}\,\Gamma_{23}}\,e^{\frac{\theta_4}{2}\,\Gamma_{34}}\,\epsilon^0_s\, ,
\end{equation}
with $\epsilon^0_s$ a constant 4-complex-component spinor; these spinors correspond to superconformal supersymmetries.
}
One can easily check that these spinors are chiral at the north pole while antichiral at the south pole, but in no way one can find a spinor (not even including the superconformal spinors) which is chiral/antichiral everywhere. Hence, in the physical theory, the Yang monopole is not supersymmetric \cite{Lambert:2014jna}.

\subsection{The topologically twisted theory: SUSY Yang monopoles}

Since we are interested in spinors which are, say, chiral on the $S^4$, it is natural to consider a 5d version of the topological twist. To that matter, we use the 't Hooft matrices to identify the $SU(2)_R$ connection with the (anti-) self-dual part of the spin connection.

\begin{equation}
(V_{a})_i^{\phantom{i}j} =\Big\{-\frac{i}{4}\,\bar{\eta}_{I\,bc}\,\omega_{abc}\,(\sigma^I)_i^{\phantom{i}j}\quad  a,b,c=1,\,2,\,3,\,4;\quad (V_{5})_i^{\phantom{i}j}=0\Big\}
\end{equation}
Then, one finds that the following spinors solve the conformal Killing spinor equation

\begin{equation}
\label{SUSY_spinors}
\epsilon^1=c\,\sqrt{r}\,\left(\begin{array}{c} 1 \\ 0 \\ 0 \\ 0 \end{array}\right)\,,\qquad \epsilon^2=c\,\sqrt{r}\,\left(\begin{array}{c} 0 \\1 \\ 0 \\ 0 \end{array}\right)\,,\qquad c \in \mathbb{C}.
\end{equation}
Imposing a reality condition on the spinor will further constrain the constant $c$. That is, depending on the choice os sign in \eqref{eq:symplectic_majorana_condition}, $c$ is an element of either $\mathbb{R}$ or $\imath\mathbb{R}$. In addition, the dilatino equation \eqref{chi_eq} is satisfied for $D=-\frac{3}{8\,r^2}$. Note that in solving this equation it is crucial that $V_{\mu}$ is independent on $r$, so that $R_{5\mu}=0$.

The spinors \eqref{SUSY_spinors}, by construction, satisfy the desired condition $\Gamma_5\,\epsilon^i=\epsilon^i$, and hence provide an unbroken supersymmetry for the Yang monopole.

Note that we could supersymmetrize instead ASD configurations involving negative chirality spinors by choosing $\Gamma_5\epsilon^i=-\epsilon^i$. The construction would have been analogous only that instead of $\bar{\eta}_{I\,ab}$ we would have needed $\eta_{I\,ab}$.

There are some subtleties arising from the behaviour of our solution at $r = 0$. Indeed, one finds that the $\sqrt{r}$ dependence of $\epsilon^i$ implies that $\epsilon^i$ is continuous yet not differentiable at $r = 0$. Similarly, $\eta^i$ has a $1/\sqrt{r}$ dependence and is thus, at least naively, singular. One can think of this behaviour in terms of a conformal transformation to $\mathbb{R} \times S^4$. The spinors \eqref{SUSY_spinors} define the conformal Killing vector $r\,\partial_r$. A conformal transformation and the coordinate change $r=e^{\tau}$ maps $\mathbb{R}^5$ into $\mathbb{R}\times S^4$. Then, the conformal Killing vector $r\,\partial_r$ becomes the actual Killing vector $\partial_{\tau}$. In turn, on $\mathbb{R}\times S^4$, it is easy to check that, upon turning on the topological twist $SU(2)_R$ gauge field, constant (and covariantly constant, hence \textit{a priori} perfectly well defined) Killing spinors, chiral on the $S^4$, can be found. From this perspective, the $r$-dependence of the spinors on $\mathbb{R}^5$ is set to $\sqrt{r}$ by the conformal mapping. 

Morover, the $SU(2)_R$ bundle has a non-vanishing second Chern class. I.e., in polar coordinates $R_i^{\phantom{i}j} \wedge R_j^{\phantom{j}i} = 3 \sin^3 \theta_1 \sin^2 \theta_2 \sin \theta_3 d\theta_1 \wedge d\theta_2 \wedge d\theta_3 \wedge d\theta_4$ and for any four-sphere surrounding the origin,

\begin{equation}
  \int_{S^4} R_i^{\phantom{i}j} \wedge R_j^{\phantom{j}i} = 8 \pi^2.
\end{equation}
$R_i^{\phantom{i}j} \wedge R_j^{\phantom{j}i}$ is closed and by Poincar\'e's lemma exact when considered on $\mathbb{R}^5$, which is clearly in contradiction with the non-trivial Chern class. By inspection one finds $R_{i}^{\phantom{i}j}$ to be singular at $r = 0$, yet not ill-defined --- intuitively one can see this in the trivial $r$-dependence of the connection. The behavior is actually that of the Yang monopole --- see \cite{Gibbons:2006wd} --- and we find that our $SU(2)_R$ connection is a Yang monopole in itself. As in the case of the Yang (and Dirac) monopole, one can deal with this behavior by either admitting singular connections or considering the theory on $\mathbb{R}^5 \setminus \{0\}$, which might be regarded as quantizing the theory in the background Yang monopole for the $SU(2)_R$ gauge field.

\section*{Acknowledgements}

The authors would like to thank Stefano Cremonesi for bringing the Yang monopole to their attention. In addition, they would like to thank Oren Bergman, Patrick Meessen and Joe Minahan for useful conversations. The authors are partly supported by the spanish grant MINECO-13-FPA2012-35043-C02-02. In addition, they acknowledge financial support from the Ramon y Cajal grant RYC-2011-07593 as well as the EU CIG grant UE-14-GT5LD2013-618459. J.S.~is supported by the Asturian government's Clar\'in FICYT grant ADC14-27.

\begin{appendix}

\section{The $SU(2)$ 5d Yang monopole}
\label{explicit_Yang}

Let us review the construction of \cite{Yang:1977qv}. To that matter, it is more convenient to write the $S^4$ metric in the $\mathbb{R}^5$ in polar coordinates in \eqref{eq_polar_coords} as an $S^3$ fibration over a disc parametrized by the polar angle $\theta\,\in\,[0,\,\pi]$ as

\begin{equation}
d\Omega_4^2=d\theta^2+\sin^2\theta\,d\Omega_3^2\, .
\end{equation}
By using the stereographic projection for the $S^3$, we write the metric of flat $\mathbb{R}^5$ as

\begin{equation}
ds^2=dr^2+r^2\,d\Omega_4^2\qquad d\Omega_4^2=d\theta^2+\frac{4\,\sin^2\theta}{(1+\vec{\xi}^2)^2}\,d\vec{\xi}^2\, .
\end{equation}

We are interested on finding self-dual or anti-self-dual  configurations on the $S^4$ with spherical symmetry. That means we need to impose $F=s\,\star_{S^4} F$ for $s=\pm 1$. In the following, for definitness, we will concentrate on $s=-1$. Setting $A_r=A_{\theta}=0$, it is straightforward to see that this requires

\begin{equation}
\label{eq:eom_A}
\partial_{\theta}\,A_i=- \frac{(1+\vec{\xi}^2)}{4\,\sin\theta}\,\epsilon^{ijk}\,F_{jk}\, ,
\end{equation}
where latin indices stand for the $\xi^i$ coordinates. Following \cite{Yang:1977qv}, we construct the matrix

\begin{equation}
B_{ij}=\frac{-4\,(1-\vec{\xi}^2)\,\delta_{ij}-8\,\xi_i\,\xi_j+8\,\epsilon_{ijk}\,\xi_k}{(1+\vec{\xi}^2)^2}\, .
\end{equation}
In terms of $B$ we can construct two vectors $\vec{b}^{\pm}$ 

\begin{equation}
b^-_i=-\frac{{\rm i}}{2}\,B_i^j\,\sigma_j\, ,\qquad b^+_i=\frac{{\rm i}}{2}\,[B^T]_i^j\,\sigma_j\,,
\end{equation}
where $\sigma^j$ are the Pauli matrices. Note that these vectors satisfy

\begin{equation}
\partial_i b^{\pm}_j-\partial_j b^{\pm}_i+[b^{\pm}_i,\,b^{\pm}_j]=0\, , \qquad [b^{\pm}_i,\,b^{\pm}_j]=\pm \frac{4}{(1+\vec{\xi}^2)}\,\,\epsilon^{ijk}\,b^{\pm}_k\, .
\end{equation}
We can now construct the gauge field  in terms of the $b^{\pm}$ and the equation it satisfies  from \eqref{eq:eom_A}:

\begin{equation}
\label{eq:eom_f}
A^{\pm}_i=f^{\pm}(\theta)\,b^{\pm}_i\, ,  \quad \partial_{\theta}f^{\pm}=\pm\frac{2}{\sin\theta}\,f^{\pm}\,(1- f^{\pm})\, .
\end{equation}
The solutions of \eqref{eq:eom_f} are

\begin{equation}
f^{\pm}=\frac{1\mp \cos\theta}{2}\, .
\end{equation}
Hence $f^{\pm}$ define, respectively, the monopole on the north and south patches of the $S^4$.

By replacing the Pauli matrices by a $N$-dimensional representation of $SU(2)$ we can embed the monopole into any $SU(N)$. It is easy to check that the instanton number \eqref{eq:instanton_number} is proportional to $N\,(N^2-1)$.

\section{Conventions}
\label{Dirac_algebra}

Our conventions are essentially those of \cite{Bergshoeff:2001hc}. The dilatino and gravitino variations of the ``standard Weyl multiplet'' are (after imposing $T = 0$)
\begin{equation}\label{eq:susy_variations}
  \begin{aligned}
    \delta \psi_\mu &= \mathcal{D}_\mu \epsilon^i - \imath \gamma_\mu \eta^i, \\
    \delta \chi^i &= \frac{1}{4} \epsilon^i D - \frac{1}{64} \gamma^{mn} R_{mn}^{ij} \epsilon_j.
  \end{aligned}
\end{equation}
We denote the Dirac matrices in tangent space as $\Gamma_a=E_{a}^\mu\,\gamma_{\mu}$, and reserve $\gamma_{\mu}$ for Dirac matrices in spacetime indices. In flat space, we choose for the former

\begin{equation}
\Gamma_{a}=\Big\{\left(\begin{array}{cc} 0 & -\,{\rm i}\,\sigma^{a} \\ {\rm i}\,\bar{\sigma}^{a} & 0\end{array}\right)\quad a=1,\,2,\,3,\,4;\qquad \Gamma_5=\Gamma_1\,\Gamma_2\,\Gamma_3\,\Gamma_4\,\Big\}\, ;
\end{equation}
where

\begin{equation}
\sigma^{a}=(\vec{\sigma},\,{\rm i}\,\unity)\,,\qquad \bar{\sigma}^{a}=(\vec{\sigma},\,-{\rm i}\,\unity)\, .
\end{equation}
On spinors such that $\Gamma_5\,\epsilon=\pm\epsilon$ and one finds that

\begin{equation}
\Gamma_5\,\epsilon=\epsilon\,\leadsto\,\Gamma_{ab}={\rm i}\,\bar{\eta}_{Iab}\,\sigma^I\, ; \qquad
\Gamma_5\,\epsilon=-\epsilon\,\leadsto\,\Gamma_{ab}={\rm i}\,\eta_{Iab}\,\sigma^I; \qquad
(a,\,b=1,\,2,\,3,\,4)\, .
\end{equation}
Here, $\eta_{Iab},\,\bar{\eta}_{Iab}$ are the 't Hooft symbols

\begin{equation}
  \eta_{Iab} = \epsilon_{Iab4} + \delta_{Ia} \delta_{b4} - \delta_{Ib} \delta_{a4}, \qquad
  \bar{\eta}_{Iab} = \epsilon_{Iab4} - \delta_{Ia} \delta_{b4} + \delta_{Ib} \delta_{a4}.
\end{equation}

$SU(2)_R$ indices are raised and lowered using the NW-SE conventions and $\epsilon^{12} = \epsilon_{12} = 1$. The covarint derivative appearing in \eqref{eq:susy_variations} are

\begin{equation}
  \mathcal{D}_\mu \epsilon^i = \nabla_\mu \epsilon^i + V_{\mu\phantom{i}j}^{\phantom{\mu}i} \epsilon^j, \qquad
  R_{\mu\nu i}^{\phantom{\mu\nu i}j} = \partial_\mu V_{\nu i}^{\phantom{\nu i}j} - \partial_\nu V_{\mu i}^{\phantom{\mu i}j} - V_{\mu i}^{\phantom{\nu i}k}V_{\nu k}^{\phantom{\nu k}j} + V_{\nu i}^{\phantom{\nu i}k} V_{\mu k}^{\phantom{\mu i}j}.
\end{equation}
Regarding the Pauli matrices, note that $\sigma^I = (\sigma^I)_i^{\phantom{i}j}$. The Wick rotation of the Lorentzian theory in \cite{Bergshoeff:2001hc} is rather straightforward with the only subtlety arising from the symplectic Majorana condition. From the gravitino variation \eqref{eq:susy_variations} it follows that $\epsilon^i$ and $\eta^i$ have to sattisfy opposite reality conditions. It seems more natural to choose the upper signs in

\begin{equation}\label{eq:symplectic_majorana_condition}
(\epsilon^i)^{\star}= \pm C\,\epsilon^j\,\epsilon_{ji}, \qquad
(\eta^i)^{\star}= \mp C\,\eta^j\,\epsilon_{ji},
\end{equation}
yet in principle the opposite sign would work just as well. The charge conjugation matrix $C$ satisfies

\begin{equation}
C\,\Gamma_a\,C^{-1}=(\Gamma_{a})^T=(\Gamma_a)^{\star}\, .
\end{equation}
When using the above basis, we pick $C = \Gamma_{13}$.

\end{appendix}

\bibliographystyle{ytphys}
\small\baselineskip=.97\baselineskip
\bibliography{ref}

\end{document}